\documentclass[aps,prl,twocolumn,superscriptaddress,groupedaddress]{revtex4}
\usepackage{subfig}
\usepackage{graphicx}  
\usepackage{dcolumn}   
\usepackage{bm}        
\usepackage{amssymb}   
\usepackage{slashed}
\usepackage{graphicx}				
\usepackage{amsmath}
\usepackage{mathtools}
\usepackage{tikz,pgf}
\usepackage{comment}
\usepackage{asymptote}
\usetikzlibrary{arrows,backgrounds}
\usetikzlibrary{fit,scopes,calc,matrix,positioning,decorations.pathmorphing}
\usepackage[all]{xy}
\usepackage{yfonts}
\newcommand{\bra}[1]{\ensuremath{\left\langle#1\right|}}
\newcommand{\ket}[1]{\ensuremath{\left|#1\right\rangle}}

\begin{document}
\title{Multipartite entanglement via the Mayer-Vietoris theorem}
\author{Andrei T. Patrascu}
\address{University College London, Department of Physics and Astronomy, London, WC1E 6BT, UK}
\address{ELI-NP, Horia Hulubei National Institute for R\&D in Physics and Nuclear Engineering, 30 Reactorului St, Bucharest-Magurele, 077125, Romania}
\begin{abstract}
The connection between entanglement and topology manifests itself in the form of the ER-EPR duality. This statement however refers to the maximally entangled states only. In this article I study the multipartite entanglement and the way in which it relates to the topological interpretation of the ER-EPR duality. The $2$ dimensional genus $1$ torus will be generalised to a $n$-dimensional general torus, where the information about the multipartite entanglement will be encoded in the higher inclusion maps of the Mayer-Vietorist sequence. 

\end{abstract}
\maketitle
\section{Introduction}
Given a 2-separable Hilbert space $\mathcal{H}=\mathcal{H}_{A}\otimes \mathcal{H}_{B}$, the wavefunction $\ket{\Psi}$ defined on it may or may not be expressible in the form $\ket{\Psi}=\ket{\psi}_{A}\otimes\ket{\psi}_{B}$. If it is, we say the state is separable, if not the state is entangled. In the case of a bipartite pure state we benefit from the singular value decomposition (Schmidt decomposition) which provides us with a local, unitary, canonical form separating the non-local parameters of the state from the trivial local ones [1,2]. This simple rule cannot be extended to multipartite entanglement [3,4]. 
Consider the set of pure states over a Hilbert space $\mathcal{H}$ which I will denote here $P(\mathcal{H})$ and the set of mixed states (convex combinations of pure states) which I will denote here as $D(\mathcal{H})=Conv(P(\mathcal{H}))$. Then the pure states will be in the set of extremal points of $D$ i.e. $P=Extr(D)$. I showed in ref. [5] that a bipartite entangled state $\ket{\psi}=\frac{1}{2}(\ket{\psi}_{1}\otimes \ket{\psi}_{2}+\ket{\psi'}_{1}\otimes \ket{\psi'}_{2})$ is induced by the inclusion maps arising in the construction of the Mayer-Vietoris theorem for the (co)homology of a 2-dimensional genus-one torus. This construction represented a preliminary mathematical proof for the ER-EPR duality [6]. Indeed, the non-trivial topology of a torus encodes an entangler gate. I arrived to this result by employing the Mayer-Vietoris long exact sequence in order to find the (co)homology of the torus from the (co)homology of the two simply connected pieces that need to be glued together to form a torus. Given the Mayer-Vietoris exact subsequence for a certain dimension, the two maps defining this subsequence induce the Hadamard and the c-NOT gates which, combined, generate entanglement. Therefore any quantum state defined with respect to a space-time having such a topology and being spread over the two ends of the ER-bridge must be entangled. That quantum states can be represented as cycles over topological spaces and can be classified by (co)homology has been shown in [5] and I will not insist on the proof here. I only wish to add that in general, quantum states may also be represented by means of higher-dimensional cycles over topological spaces, while keeping the same topological properties and algebraic operations that are valid for 1-dimensional cycles. This observation allows us to employ the long Mayer-Vietoris sequence and to focus on higher order subsequences thereof, while discussing multipartite entanglement [7-9]. While the proof of the ER-EPR conjecture for bipartite entanglement is now better understood, the connection between multipartite entanglement and topologically non-trivial spaces presents a completely new level of complexity. The mystery lies on both sides: on the EPR side, multipartite quantum entanglement is still not completely understood [10]. The lack of a generalisation of or a replacement for the Schmidt decomposition for multipartite entanglement still slows the development of quantum computing. As maximal bipartite entanglement cannot be shared over more than two subsystems [11] we need a better understanding of  multipartite entanglement. A tool for better understanding this may prove to be the Mayer-Vietoris theorem. 
\par The topological invariants of a torus given by homology and cohomology can be derived by means of sheets covering subspaces of the torus, usually of lower dimension. Together, such sheets are capable of encoding the global information within the torus and therefore become an indicator for the non-local entanglement effects. 

From an operational point of view, the main idea is to cover the two handles forming a torus by two lower dimensional sheets. One defines the quantum states on the sheets. The quantum states are unaffected by continuous deformations of the sheets. These deformations however, may lead to the intersection of the sheets associated to the two handles. The maps generating the Mayer-Vietoris sequence for these sheets induce quantum multipartite entangler gates as follows: the inclusion inserting the intersections into the respective sheets forms a map which, given a twisted torus, is precisely the Hadamard gate. Together with the other map arising in the Mayer-Vietoris sequence which can be regarded as a c-NOT gate, we obtain a sequence that has the capability of inducing quantum bipartite entanglement. Generalising this discussion for higher genus surfaces and higher dimensions leads to long exact Mayer-Vietoris sequences and a hierarchy of maps inducing multipartite entanglement. It is worth mentioning that entanglement arises from the maps that include subspaces into the larger topological spaces with non-trivial global structure. 
As shown in [5], a qubit can be represented by means of a worldline. Any qubit can be written as a superposition of states by means of the Hadamard matrix. The resulting states can then be transported along different spacetime trajectories. In topological terms, this implies that any qubit can be represented by means of a topological cycle. Such a cycle may be trivial i.e. it may be reduced back to a worldline, meaning the topology of spacetime is trivial and the quantum states are not entangled. Alternatively, the cycle may not be trivial i.e. it may not be possible to reduce it to a worldline. The only reason for a cycle to be non-reducible is the existence of a topological obstruction. This appears to be the case when our states have been transported across an ER bridge. This observation strengthens the argument in favour of EPR$\Rightarrow$ER.

The relevant section of the Mayer-Vietoris sequence for the bipartite system is
\begin{widetext}
\begin{equation}
...\rightarrow H_{2}(A)\oplus H_{2}(B)\rightarrow H_{2}(T^{2})\xrightarrow{\partial} H_{1}(A\cap B)\xrightarrow{(i_{*},j_{*})}H_{1}(A)\oplus H_{1}(B)\rightarrow ... 
\end{equation}
\end{widetext}
where $A$ and $B$ are the respective sheets partitioning the surface of the torus into two covers of the respective sides. 
Replacing the known homologies one arrives at 

\begin{equation}
...\rightarrow 0\rightarrow H_{2}(T^{2})\xrightarrow{\partial}\mathbb{Z}\oplus\mathbb{Z}\xrightarrow{(i_{*},j_{*})}\mathbb{Z}\oplus\mathbb{Z}\rightarrow ... 
\end{equation}
I proved in ref. [5] that for a twisted torus the map $(i_{*},j_{*})$ connecting the homology of the intersection of the two sheets $A$ and $B$ to the sum of the two distinct homology groups of the sheets $A$ and $B$ has the precise form 
\begin{equation}
\left( \begin{array}{cc}
1 & 1  \\
1 & -1  \\
 \end{array} \right):\mathbb{Z}\oplus\mathbb{Z}\rightarrow \mathbb{Z}\oplus\mathbb{Z}
 \end{equation}
 which is precisely the Hadamard matrix. The remaining map in the sequence is the c-NOT map. When these two are taken together to detect the (co)homology of a torus we obtain the standard entangler gate. 

\section{Multipartite entanglement and separability}
An important technical aspect employed in this article is that bipartite and multipartite quantum entanglement can be represented by means of the maps appearing in the Mayer-Vietoris theorem. At this point I show how bipartite and multipartite entanglement appear as consequences of the spacetime topology and what types of multipartite entanglement are possible. I also show that the maps resulting from the long exact Mayer-Vietoris sequence can be seen as entanglement maps in the more general higher dimensional cases and also for higher genera. 
It is important to note that the no-cloning theorem which leads to the impossibility of sending faster than light signals via quantum entanglement remains valid on the gravity side. Cloning would imply non-unitary creation of entanglement. On the gravity side, it has been shown in [13] that this would imply an additional change in spacetime topology not allowed both by topology and entanglement conservation theorems. As has been noticed in [13], the topology conservation theorems do not rule out processes like black hole pair production in the context of ER-EPR. The creation of a pair of entangled black holes does not change the topology as the ER bridge between them is formed from the Planckian wormholes connecting the entangled vacuum. Therefore, in a sense, the non-trivial topology was already there. Interpreting the standard quantum mechanical entanglement (say, between electrons) as a nontrivial topology may be possible provided we look at the topological properties of maps between spacetime patches. This would imply the use of a categorial language, a topic I will leave for a future article. 
For the multipartite entanglement case, the main topological object will be a higher genus surface which will, as before, not violate the no-cloning theorem. 
In order to continue I briefly introduce first the notation following ref. [10]. 
In general a state vector is an element of a specific Hilbert space $\ket{\psi}\in \mathcal{H}$. We assume it to be normalised as 
\begin{equation}
||\psi||=\sqrt{\bra{\psi}\ket{\psi}}=1
\end{equation}
A pure state is generally defined as a one dimensional subspace (ray) of the Hilbert space which can be given by a state vector as the self-adjoint linear operator $\pi=\ket{\psi}\bra{\psi}$ projecting to the one dimensional subspace spanned by the state vector $\ket{\psi}$ where $\pi^{2}=\pi=\pi^{\dagger}$. The set of pure states over the Hilbert space is 
\begin{equation}
P(\mathcal{H})=\{\pi\in Lin_{SA}\mathcal{H}|\pi^{2}=\pi,||\pi||_{tr}=tr(\pi)=1\}
\end{equation}
A mixed state is represented by the convex combination of pure states. It can be seen as the ensemble of quantum systems $\{(p_{i},\pi_{i})|i=1,...,m\}$ with the pure states $\pi_{i}$ occurring with the probability $p_{i}$. The Hilbert space $\mathcal{H}$ contains the set of mixed states defined as 
\begin{widetext}
\begin{equation}
D(\mathcal{H})=Conv(P(\mathcal{H}))=\{\rho\in Lin_{SA}(\mathcal{H})| \exists \pi_{i}\in P, p_{i}\geq 0,\sum\limits_{i}p_{i}=1:\rho=\sum\limits_{i}p_{i}\pi_{i}\}
\end{equation}
\end{widetext}
Looking at the states of the Hilbert space geometrically one may say that the pure states are the extremal points of the convex polytope defined by the mixed states i.e. 
\begin{equation}
P=Extr\{D\}
\end{equation}
In order to speak about entanglement however we need composite quantum systems. We start with a Hilbert space which can be written in terms of the direct product of two subsystems 
\begin{equation}
\mathcal{H}_{12}=\mathcal{H}_{1}\otimes \mathcal{H}_{2}
\end{equation}
We can define pure states on the two subsystems as $P_{1}=P(\mathcal{H}_{1})$ and $P_{2}=P(\mathcal{H}_{2})$. The pure states on the product space will be $P_{12}=P(\mathcal{H}_{12})$. Analogously we have for the mixed states $D_{1}=D(\mathcal{H}_{1})=Conv(P_{1})$, $D_{2}=D(\mathcal{H}_{2})=Conv(P_{2})$ and $D_{12}=D(\mathcal{H}_{12})=Conv(P_{12})$.The marginal states (or reduced states) are obtained by employing the partial trace i.e. $tr_{2}=D_{12}\rightarrow D_{1}$. If the state vector $\ket{\psi}\in \mathcal{H}_{12}$ can be written in terms of a direct product of vectors on each of the Hilbert subspaces as in $\ket{\psi}=\ket{\psi_{1}}\otimes \ket{\psi_{2}}$ then we say it is separable. If this is impossible we have an entangled state. The set of separable pure states can be written as
\begin{equation}
P_{sep}=\{\pi\in P_{12}| \exists \pi_{1}\in P_{1},\exists \pi_{2}\in P_{2}: \pi=\pi_{1}\otimes \pi_{2}\}
\end{equation}
The set of entangled pure states is its complement $\bar{P}_{sep}=P-P_{sep}$. The set of separable mixed states is $D_{sep}=Conv(P_{sep})$ while the set of entangled mixed states is its complement $\bar{D}_{sep}=D-D_{sep}$. It is worthwhile to notice that separable mixed states can be created from uncorrelated product states by the use of local operations and classical communication. Entangled states cannot be created by such methods. However, starting with entangled states one can obtain separable states by means of local operations and classical communication. 
\par For multipartite states the situation is different. In order to discuss it let me first introduce a few notations. Let the labels of the respective elementary subsystems be $L=\{1,2,...,n\}$. Given one such label $a\in L$ we have the Hilbert space $\mathcal{H}_{a}$ associated to the elementary subsystem of label $a$. A non-elementary subsystem can be labeled by a subset of such labels, denoted $K\subseteq L$. The associated Hilbert space is then 
\begin{equation}
\mathcal{H}_{K}=\bigotimes\limits_{a\in K}\mathcal{H}_{a}
\end{equation}
For a subsystem $K$ we have the set of pure states 
\begin{equation}
P_{K}=P(\mathcal{H}_{K})
\end{equation}
and the set of mixed states
\begin{equation}
D_{K}=Conv(P_{K})
\end{equation}
For pure and mixed states of the whole subsystem I employ the notations $P=P_{L}$ and $D=D_{L}$. 
A given kind of partially separable pure states can be used in order to form mixed states. Consider for example 
\begin{equation}
\alpha=\{K_{1}, K_{2}, ..., K_{|\alpha|}\}=K_{1}|K_{2}|...|K_{|\alpha|}
\end{equation}
This represents a splitting of the system i.e. a partition of the labels $L$ into parts. The set of all possible partitions is
\begin{widetext}
\begin{equation}
P_{I}=\{\alpha=K_{1}|K_{2}|...|K_{|\alpha|}|\forall K\in \alpha: K\in P_{0}-\{\emptyset\}, \forall K, K' \in \alpha : K\neq K'\Rightarrow K\cap K'=\{\emptyset\}, \bigcup\limits_{K\in \alpha}K=L\}
\end{equation}
\end{widetext}
These partitions are called labels of first kind. We can also define a partial order on the partitions. Given two partitions $\alpha$ and $\beta$ $\in P_{I}$, $\beta$ is a refinement of $\alpha$ if $\alpha$ can be obtained from $\beta$ by joining some of the parts of $\beta$
\begin{equation}
\beta \preceq \alpha \xLeftrightarrow{def} \forall K' \in \beta, \exists K \in \alpha : K'\subseteq K
\end{equation}
The set $(P_{I},\preceq)$ is a poset. 
One can define the minimal and maximal elements of a subset $Q\subseteq P$ as 
\begin{equation}
\begin{array}{c}
min(Q)=\{x\in Q| (y\in Q,\, y \preceq x)\Rightarrow y=x\}\\
\\
max(Q)=\{x\in Q| (y\in Q,\, x\preceq y)\Rightarrow y=x\}\\
\\
\end{array}
\end{equation}
A subset $Q\subseteq P$ is a down-set (also called order ideal) if
\begin{equation}
(x\in Q,\, y\preceq x)\,\Rightarrow \, y\in Q
\end{equation}
The set of all up-sets of $P$ is denoted with $\mathcal{O}_{\uparrow}(P)$. For a subset $Q\subset P$ one can define 
\begin{equation}
\begin{array}{c}
\downarrow Q =\{x\in P|\exists y\in Q : x\preceq y\}\\
\\
\uparrow Q=\{x\in P|\exists y\in Q : y\preceq x\}\\
\\
\end{array}
\end{equation}
called the down-set and the up-set respectively. 
One can define for a poset the greatest lower bound or meet $\alpha\wedge\alpha'$ and the least upper bound or join $\alpha\vee \alpha'$ as 
\begin{equation}
\begin{array}{c}
\alpha\wedge \alpha'=\{K\cap K'\neq \emptyset | K \in \alpha, K'\in \alpha '\}\\
\\
\alpha\vee \alpha'=\bigwedge \uparrow\{\alpha,\alpha'\}\\
\end{array}
\end{equation}
One may notice that bipartition can be used in order to generate all other partitions 
\begin{equation}
\alpha=\bigwedge\limits_{K\in\alpha} K|\bar{K}
\end{equation}
Given a partition $\alpha\in P_{I}$ we have the set of $\alpha$-separable pure states 
\begin{equation}
P_{\alpha}=\{\pi\in Lin(\mathcal{H})|\forall K\in \alpha, \exists \pi_{K}\in P_{K}:\pi=\bigotimes\limits_{K\in \alpha}\pi_{K}\}
\end{equation}
and the set of $\alpha$-separable mixed states
\begin{equation}
D_{\alpha}=Conv(P_{\alpha})
\end{equation}
One can say that a state is $\alpha$-separable if and only if it can be mixed by the use of $\alpha$-separable pure states. 
A pure state $\pi\in P$ is $\alpha$-entangled if it is not $\alpha$-separable. $\bar{P}_{\alpha}$ is not closed. A mixed state is $\alpha$-entangled if it is not $\alpha$ separable. One also has to notice that there exist mixed states which cannot be mixed by using any given non-trivial $\alpha$-separable pure states while they can be mixed by the use of pure states of different non-trivial $\alpha$-separability. For example for the tripartite case $\mathcal{H}_{123}=\mathcal{H}_{1}\otimes \mathcal{H}_{2}\otimes \mathcal{H}_{3}$ the sets of $\alpha$-separable pure states can be written as 
\begin{equation}
\begin{array}{c}
P_{123}=P(\mathcal{H}_{123})\\
\\
P_{a|bc}=\{\pi\in P_{123}| \pi=\pi_{a}\otimes \pi_{bc}\}\\
\\
P_{1|2|3}=\{\pi\in P_{123}|\pi=\pi_{1}\otimes \pi_{2}\otimes \pi_{3}\}\\
\\
\end{array}
\end{equation}
The sets of $\alpha$-separable mixed states are 
\begin{equation}
\begin{array}{c}
D_{123}=Conv(P_{123})=D(\mathcal{H}_{123})\\
\\
D_{a|bc}=Conv(P_{a|bc})\\
\\
D_{1|2|3}=Conv(P_{1|2|3})\\
\end{array}
\end{equation}
There are states $\rho\notin D_{a|bc}$ which can be mixed by the use of bipartite entanglement in subsystems $12$, $13$ and $23$ i.e. $\rho \in Conv(P_{1|23}\cup P_{2|13}\cup P_{3|12})$. States of this type cannot be considered as fully tripartite entangled since they can be mixed without the use of genuine tripartite entanglement. Therefore, mixtures from different types of partially separable pure states can also be considered. 
Let $\bar{\alpha}$ be a nonempty down-set in $P_{I}$, i.e. $\bar{\alpha}=\{\alpha_{1}, \alpha_{2}, ..., \alpha_{|\bar{\alpha}|}\}\subseteq P_{I}$ containing every partition which is finer than its maximal elements. The set of all possible non-empty down-sets is 
\begin{equation}
\begin{array}{c}
P_{II}=\mathcal{O}_{\downarrow}(P_{I})-{\emptyset}=\\
\\
=\{\bar{\alpha}\in 2^{P_{I}}-\{\emptyset\}|\forall \alpha\in\bar{\alpha}:\beta\preceq\alpha \Rightarrow \beta\in\bar{\alpha}\}
\\
\end{array}
\end{equation}
The non-empty down-sets of partitions are called labels of second kind. They are used to label states with the respective partial separability.
The set of non-empty down-sets $\bar{\alpha}$ determines $\bar{\alpha}\downarrow max(\bar{\alpha})$ uniquely we can use either $max(\bar{\alpha})$ or $\bar{\alpha}$ for labelling of sets of states with the specific partial separability properties. 
For a down-set $\bar{\alpha}\in P_{II}$, we have the set of $\bar{\alpha}$-separable pure states
\begin{equation}
P_{\bar{\alpha}}=\bigcup\limits_{\alpha\in \bar{\alpha}}P_{\alpha}=\bigcup\limits_{\alpha \in max(\bar{\alpha})}P_{\alpha}
\end{equation}
This means that $\pi$ is $\bar{\alpha}$-separable iff it is $\alpha$-separable for at least one $\alpha\in\bar{\alpha}$.
The set of $\bar{\alpha}$-separable mixed states is
\begin{equation}
D_{\alpha}=Conv(P_{\alpha})
\end{equation}
This means that a state $\rho$ is $\bar{\alpha}$-separable iff it can be mixed by the use of any $\alpha$-separable pure states with $\alpha\in\bar{\alpha}$. Therefore 
\begin{equation}
P_{\bar{\alpha}}=Extr(D_{\bar{\alpha}})
\end{equation}
In the language of ref. [5] separability of a certain order is to be associated with the possibility of having distinct chains over sheets covering a topological space. When such chains are independent they are described by a certain (co)homology group. When going to higher dimensions, such chains may join and form either topologically trivial objects (described by trivial (co)homology groups) case in which we have higher-dimensional separability, or they can join around holes in our topological space, forming topologically non-trivial objects characterised by non-trivial higher (co)homology groups. When all these sheets are immersed the overall topological space, each subsequence of the long Mayer-Vietoris sequence will encode a certain type of entanglement, visible given the dimension specific to the subsequence.  This will produce a full hierarch of multipartite entanglement when all subsequences are considered together. 

The choice of the covers will now be more complicated as one has to cover several handles that will eventually come together to form a higher genus torus. This will imply a combinatorial hierarchy of multipartite entanglement, as verified also by [10]. As a basic example, in the case of a genus $3$ torus a one dimensional cycle will be able to encircle one hole in the torus, this being described as entanglement via a lower subsequence of the long Mayer-Vietoris sequence, and only then glue via another Mayer-Vietoris subsequence to a chain coming from the other side of the torus. This will then lead to a set of potential combinations of chains and cycles that may cover our torus. Similarly, two 2-dimensional sheet can encircle each one hole of the torus and then glue together via a higher Mayer-Vietoris subsequence, somewhere in the middle. All possible examples can be seen in Fig. 2 

\section{Higher subsequences of Mayer-Vietoris}
\par At this point we have all the notations required for the development and proof of the new results. 
Now that the definition of bipartite and multipartite entanglement is clear and we understand that multipartite entanglement is of several types, I present the arguments connecting quantum entanglement to the maps in the Mayer-Vietoris sequence of algebraic topology. Indeed, figure 2 shows the labelling in the case of $n=3$ states and their possible entanglement. Indeed, all possible outcomes can be described in terms of a genus $3$ torus as depicted in the figure. Heuristically, let me now define on the torus subspaces of dimension one (i.e. curves) such that each of the covers may turn around one hole in the torus once without turning around the other. This represents the lowest graph in figure 2 on the left. As we select sheets turning around two holes in the torus we arrive at the set of three graphs in the middle and, finally, covering the whole genus 3 torus leads us to the upper graph. This represents $P_{I}$ in the notation above. However, these are not the sole constructions resulting from the Mayer-Vietoris sequence. Indeed, one can go to higher dimensions and define covers as surfaces on the torus. The different ways in which they can overlap will define the higher separability and entanglement. For example labels of type $P_{II}$ can be defined by means of two types of coverings intersecting on the genus three torus. It is also possible to introduce entanglement by intersecting covers of one dimension with covers of another dimension. The distinction between the case when sheets of the same dimension intersect and when sheets of different dimensions intersect can be seen as the distinction between mixtures from identical types of separable pure states and mixtures from different types. In what follows I closely follow ref. [5] and present first bipartite entanglement. I then show the power of the full Mayer-Vietoris sequence for detecting the entanglement maps in a genus 3 torus and how they connect to the $\alpha$-entanglement. 

Let me therefore start with the short Mayer Vietoris sequence 
\begin{widetext}
\begin{equation}
... \rightarrow H_{2}\oplus H_{2}(B)\rightarrow H_{2}(T^{2})\xrightarrow{\partial}H_{1}(A\cap B)\xrightarrow{(i_{*},j_{*})} H_{1}(A)\oplus H_{1}(B)\xrightarrow ...
\end{equation}
\end{widetext}
Here $T^{2}$ is the standard two dimensional, genus one torus. $A$ and $B$ represent partial cover sheets for the two handles forming the torus. The index of the homology groups represent the dimension for which we calculate the homology. It is important to notice that, except for the homology of the actual torus, in this sequence we can calculate each of the other homology groups. Indeed, this simplification occurs due to the particular choice of the cover sheets. The sequence above therefore becomes, after replacing the known homology groups with their respective values 
\begin{equation}
... \rightarrow 0 \rightarrow H_{2}(T^{2})\xrightarrow{\partial}\mathbb{Z}\oplus \mathbb{Z}\xrightarrow{(i_{*},j_{*})}\mathbb{Z}\oplus\mathbb{Z}\rightarrow ...
\end{equation}
It is important to notice that while the two groups appearing in the sequence on the right are isomorphic (obviously $\mathbb{Z}\oplus \mathbb{Z}\cong \mathbb{Z}\oplus \mathbb{Z}$) the map between them $(i_{*}, j_{*})$ does not have to be an isomorphism. In fact it cannot be one because it originates from the Mayer-Vietoris map 
\begin{equation}
H_{1}(A\cap B)\xrightarrow{(i_{*},j_{*})} H_{1}(A)\oplus H_{1}(B)
\end{equation}
 which includes the intersection in the two covering sheets defined on the two handles of the torus. If we take $1$-cycles generating the homologies of $A$, $B$ and $A\cap B$ such that for each cylinder formed by the intersection $A\cap B$ the cycle is the equatorial circumference and the associated homology classes are $\alpha$ and $\beta$, then these cycles each generate a free abelian group $\mathbb{Z}$ with the inclusion 
\begin{equation}
(i_{*},j_{*}):\mathbb{Z}[\alpha]\oplus \mathbb{Z}[\beta]\hookrightarrow \mathbb{Z}[\alpha]\oplus\mathbb{Z}[\beta]
\end{equation}
but $\alpha=\beta$ when we are on the sides of $H_{n}(A)$ and $H_{n}(B)$ and therefore 
\begin{equation}
(i_{*},j_{*})(\alpha,0)=(i_{*},j_{*})(0,\beta)=(\alpha,\beta)
\end{equation}
The map  connecting the two rightmost groups is then 
\begin{equation}
 \left( \begin{array}{cc}
1 & 1  \\
1 & 1  \\
 \end{array} \right) : \mathbb{Z}\oplus\mathbb{Z}\rightarrow \mathbb{Z}\oplus\mathbb{Z}
\end{equation}
and for a twisted torus, the map becomes 
\begin{equation}
 \left( \begin{array}{cc}
1 & 1  \\
1 & -1  \\
 \end{array} \right) : \mathbb{Z}\oplus\mathbb{Z}\rightarrow \mathbb{Z}\oplus\mathbb{Z}
\end{equation}
But this map arising solely from considerations intrinsic to the Mayer-Vietoris sequence is precisely the Hadamard map. Together with the other map of the current subsequence of the Mayer-Vietoris sequence, this will generate an entangler gate.

\begin{figure*}
\centering
\includegraphics[width=130pt]{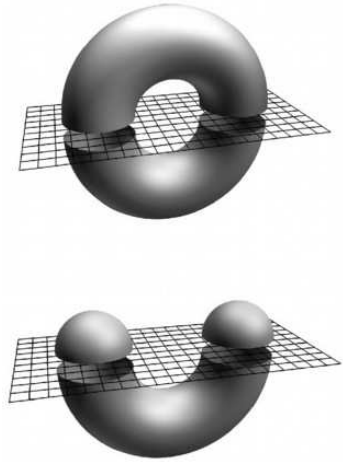}
\caption{Cutting the surface and assigning sheets to each of the resulting surfaces, then measuring the inclusion of the intersection of the surfaces in the respective sheets provides global information about the surface. One can therefore distinguish a torus from a deformed sphere by the fact that for a torus, the inclusion cannot be reduced back to a sphere by any smooth deformation.
This is a basic example for the $2$-dimensional torus. Cutting the torus by a planar surface gives two regions. Covering both of them with corresponding sheets and allowing those sheets to intersect on the handles of the torus will provide us with information about the topology of the torus [12]. Moreover, this construction also inevitably leads to inclusion maps of the form of Hadamard and c-NOT operations leading to an entangler gate.
}
\label{fig:top}
\end{figure*}

\begin{figure*}
\centering
\includegraphics[width=230pt]{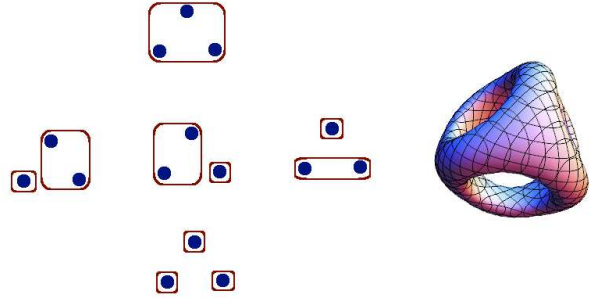}
\caption{Different types of multipartite entanglements and the higher genus toroidal representation}
\end{figure*}

At this moment we have a proof for the connection between bipartite entanglement and genus-one, two-dimensional toroidal topology. In what follows I show that a similar argument can be brought for multipartite entanglement. 
The main tool in what follows will be the long exact Mayer-Vietoris sequence
\begin{widetext}
\begin{equation}
\begin{array}{c}
...\rightarrow H_{n+1}(X)\xrightarrow{\partial_{*}}H_{n}(A\cap B)\xrightarrow{(i_{*},j_{*})}H_{n}(A)\oplus H_{n}(B)\xrightarrow{k_{*}-l_{*}}H_{n}(X)\xrightarrow{\partial_{*}} H_{n-1}(A\cap B)\rightarrow ... \\
\\
...\rightarrow H_{0}(A)\oplus H_{0}(B)\xrightarrow{k_{*}-l_{*}}H_{0}(X)\rightarrow 0\\
\end{array}
\end{equation}
\end{widetext}
I showed in ref. [5] that the topology of the underlying space (for example spacetime) can be used as an indicator for bipartite entanglement. Indeed, as stated by the ER-EPR duality, topological wormholes are equivalent to quantum entanglement. Physical observables are to be associated to the cohomology groups and entanglement can be detected by means of the maps between the various (co)homology groups appearing in the Mayer-Vietoris sequence. It therefore makes sense to extend the application of Mayer-Vietoris sequence in cohomology for higher dimensional tori. 
The first step in employing Mayer-Vietoris to describe entanglement is to calculate the cohomology of the n-dimensional torus $X=(S^{1})^{n}\cong \mathbb{R}^{n}/\mathbb{Z}^{n}$. This can be done inductively by looking at this torus as a union of two copies of $I\times X^{n-1}$. These copies intersect as the disjoint union $X^{n-1}\sqcup X^{n-1}$. Replacing this in the Mayer-Vietoris sequence we obtain 
\begin{widetext}
\begin{equation}
... \rightarrow H^{k}(X^{n})\xrightarrow{a_{k}}H^{k}(X^{n-1})\oplus H^{k}(X^{n-1})\xrightarrow{b_{k}}H^{k}(X^{n-1}\sqcup X^{n-1})\xrightarrow{d_{*}}H^{k+1}(X^{n})\rightarrow ...
\end{equation}
\end{widetext}
The particular maps are as follows: the maps $a_{k}$ are induced by the inclusion
\begin{equation}
(I\times X^{n-1})\sqcup(I\times X^{n-1})\hookrightarrow S^{1}\times X^{n-1}
\end{equation}
They connect the cohomology of $X^{n}$ to the cohomology of $X^{n-1}$ and therefore account for the decrease in the dimension of the subspace considered in the Mayer-Vietoris sequence. The analogy with quantum multipartite entanglement is obvious: states of multipartite entanglement can be separable with respect to a bipartition containing on one side one single element and on the other side the rest of $(n-1)$ elements. 
The maps $b_{k}$ are induced by the difference between the inclusions 
\begin{equation}
\{0\}\times X^{n-1}\sqcup \{1\}\times X^{n-1}\hookrightarrow I\times X^{n-1}
\end{equation}
These maps are homotopic to $Id:X^{n-1}\rightarrow X^{n-1}$. We can think of $b_{k}$ as being given by 
\begin{equation}
a\oplus b \rightarrow (a-b)\oplus (b-a)
\end{equation}
with $a,b\in H^{k}(X^{n-1})$ and $a_{k}$ is given by two copies of a single map 
\begin{equation}
H^{k}(X^{n})\xrightarrow{a'_{k}} H^{k}(X^{n-1})
\end{equation}
This shows that the same separability level can be obtained by mixing states of different $\alpha$-separability. 
This leads us to the conclusion that 
\begin{equation}
Im(b_{k})=Ker(d_{k})
\end{equation}
is the anti-diagonal copy $a\oplus (-a)$ of $H^{k}(X^{n-1})$ in $H^{k}(X^{n-1})\oplus H^{k}(X^{n-1})$.
Therefore the map $d_{k}$ simply factors through the map $a\oplus b\rightarrow (a+b)$ connecting $H^{k}(X^{n-1})\oplus H^{k}(X^{n-1})$ to $H^{k}(X^{n-1})$ and $H^{k+1}(X^{n})$ contains a copy of $H^{k}(X^{n-1})$ given by $Im(d_{k})$. Also 
\begin{equation}
Im(a_{k})=Ker(b_{k})
\end{equation}
 is a diagonal copy $a\oplus a$ of $H^{k}(X^{n-1})\oplus H^{k}(X^{n-1})$ and therefore $a'_{k}:H^{k}(X^{n})\rightarrow H^{k}(X^{n-1})$ is surjective. Finally, 
\begin{equation}
Ker(a_{k+1})=Im(d_{k})
\end{equation}
is the kernel of $a'_{k+1}$ and hence the quotient $H^{k+1}(X^{n})/Im(d_{k})$ is precisely $H^{k+1}(X^{n-1})$. Therefore we obtain the short exact sequence 
\begin{equation}
0\rightarrow H^{k}(X^{n-1})\rightarrow H^{k+1}(X^{n})\rightarrow H^{k+1}(X^{n-1})\rightarrow 0
\end{equation}
By induction, the Poincare polynomials 
\begin{equation}
p_{n}(t)=\sum\limits_{k=0}^{n}dim(H^{k}(X^{n}))t^{k}
\end{equation}
satisfy the relation $p_{0}(t)=1$ and $p_{n}(t)=(1+t)p_{n-1}(t)$ and therefore 
\begin{equation}
p_{n}(t)=(1+t)^{n}=\sum\limits_{k=0}^{n} \binom nk t^{k}
\end{equation}
By induction we conclude that $H^{k}(X^{n})\cong \mathbb{Z}^{\binom nk}$
\par This calculation has shown how higher dimensions of the higher genus torus affect the maps within the Mayer Vietoris sequence. Quantum mechanically, changing the dimension of the torus leads us to maps arising between homologies given by combinatorial powers of abelian groups. In order to change the order of the entanglement it is sufficient to increase the connectivity of the space i.e. to increase the genus of the torus. This can be achieved by puncturing the torus and gluing it with another torus over the cut. This procedure can be repeated leading to higher genus tori 
\begin{equation}
T_{g}=T^{2}\wedge T^{2}\wedge ... \wedge T^{2}
\end{equation}
where the wedge operation occurs $g$ times. The quantum entanglement analogue to this statement is that bipartition can be used to generate all other partitions and the associate entanglement has at its coarsest form a bipartite entanglement between two subsystems, in this case two genus 1 tori. 


In the multipartite case there will be more inclusion maps. Also, higher dimensional inclusion maps will emerge. The basic idea is to consider multipartite entanglement as given by all the standard inclusion maps arising in the long Mayer-Vietoris sequence describing a higher dimensional torus. The combinatorial nature of these maps will produce the hierarchy of multipartite entanglement. 
\begin{widetext}
\begin{equation}
\begin{array}{cc}
...\rightarrow H_{n+1}(X)\xrightarrow{\partial_{*}}H_{n}(A\cap B)\xrightarrow{(i_{*},j_{*})}H_{n}(A)\oplus H_{n}(B)\xrightarrow{k_{*}-l_{*}}H_{n}(X)\xrightarrow{\partial_{*}}H_{n-1}(A\cap B)\rightarrow \\
\\
...\rightarrow H_{0}(A)\oplus H_{0}(B)\xrightarrow{k_{*}-l_{*}}H_{0}(X)\rightarrow 0\\
\\
\end{array}
\end{equation}
\end{widetext}

Each of the maps has a dimension equal to the combinations that can be formed taking the dimension of the subspace considered in the respective region of the long Mayer-Vietoris sequence and the degree of the (co)homology in that region. Each section of Mayer-Vietoris sequence will therefore encode a certain $\alpha$-separability and an associated $\alpha$-entanglement [10]. 
Moreover, as I have shown for the genus 3 torus, a generalisation to higher genus tori takes into account more of the complexity given by multipartite entanglement. Such a higher genus extension can be achieved by puncturing the torus and gluing it with another torus over the cut. This procedure can be repeated leading to higher genus tori. For genus $g$ we have
\begin{equation}
T_{g}=T^{2}\wedge T^{2}\wedge ... \wedge T^{2}
\end{equation}
where the wedge operation occurs $g$ times. The quantum entanglement analogue to this statement is that bipartition can be used to generate all other partitions and the associated entanglement has at its coarsest form a bipartite entanglement between two subsystems, in this case two genus 1 tori. As showed in ref. [5] the quantum states belong to the covers used for the construction of the Mayer-Vietoris sequence for the general oriented surface. As such, they form (co)homology classes and are therefore computable by means of homological algebraic tools. A pure state $\ket{\Psi_{n,n}}$ is said to describe an $n$-partite quantum system and to be $n$-partite entangled if for any separation of the system in two subsystems the state $\ket{\Psi_{n,n}}$ cannot be described as a tensor product of states of these two subsystems. An $n$-partite state $\ket{\Psi_{k,n}}$ is said to be $k$-partite entangled if it cannot be described without at least one $k$-partite entangled sub-state. A mixed $n$-partite state is considered $k$-partite entangled if it cannot be expressed as a statistical mixture
\begin{equation}
\sum\limits_{i} \sum\limits_{j=1}^{k-1} p_{ij} \ket{\Psi^{(i)}_{j,n}}\bra{\Psi_{j,n}^{(i)}}\neq \rho_{k,n}
\end{equation}
with at most $(k-1)$-partite entangled states with $p_{ij}\geq 0$. 
This shows that the structure of multipartite entanglement is non-trivial depending on the different combinations accessible to the subsystems. 
To underline how multipartite entanglement is equivalent to higher genus surfaces patched together via the maps of the Mayer-Vietoris sequence let me consider the tripartite entanglement for the so called $W$-state 
\begin{equation}
\ket{W}=\frac{1}{\sqrt{3}}(\ket{001}+\ket{010}+\ket{100})
\end{equation}
describing the entanglement of three qubits such that when the entanglement of one of the three qubits is lost, the state of the remaining 2-qubit system is still entangled.
Let me also consider the $GHZ$-state
\begin{equation}
\ket{GHZ}=\frac{1}{\sqrt{2}}(\ket{0}^{\otimes 3}+\ket{1}^{\otimes 3})
\end{equation}
which loses the whole entanglement whenever the entanglement of one of the three qubits is lost. To see how this leads to the genus 3 torus depicted in figure 2 let us start with the situation in which we have full tripartite entanglement for both $\ket{W}$ and $\ket{GHZ}$. This is represented by the top left element of figure 2. In terms of coverings of the genus 3 torus, this means we take into account the surface generating the whole torus with all subspaces included and therefore representing the final subchain of the Mayer-Vietoris sequence. Moving in the long Mayer-Vietoris sequence to the left we already obtain two options: either we overlap the subsurface obtained by covering the genus 2 torus with the subsurface of trivial topology extending the trivial surface to generate the three torus, case in which the maps will generate the $\ket{W}$ state which is robust to particle loss, in the sense that the original states would remain bipartite entangled, or we take trivial subsurfaces on each vertex of the genus 3 torus which we may as well unite into a three-partite entangled state but which would lose all entanglement once the entanglement of one qubit is lost. The first situation is encoded in the second row of figure 2 while the last situation is encoded in the lower part of figure 2.

\section{Conclusions}
 As stated by the ER-EPR duality, topological wormholes are equivalent to quantum entanglement. Physical observables are to be associated to the cohomology groups [5] and entanglement can be detected by means of the maps between the various (co)homology groups appearing in the Mayer-Vietoris sequence. It therefore makes sense to extend the application of Mayer-Vietoris sequence in cohomology for higher dimensional and arbitrary genus tori. 
 
The classical limit of the ER-EPR duality, involving general relativity on the gravity side has been discussed in [14]. Understanding the gravitational side of the duality when the entanglement involves standard model particles like electrons or photons represents a major challenge. The fact that the maps in the Mayer-Vietoris sequence reproduce entangling gates may indicate a new way of looking at this challenge. Indeed, what if the quantum gravity side of the entanglement of photons would correspond not to properties of spaces but instead to properties of the classes of maps connecting patches of spaces together into globally meaningful structures? This idea may lead to a category theory approach to quantum gravity where maps between spacetime patches may play a more important role when trying to understand quantum gravity effects. This idea is the subject of future research.

It is particularly relevant to note that analysing the higher genus torus by means of the Mayer-Vietoris sequence we obtained the expected (co)homology structure which depends on the combinations of the level of the partitions of a system $(n)$ by the level of the entangled subsystems $(k)$. Such a formula introducing the combinations of the subdivisions by the level of entanglement is compatible with other combinatorial tools for identifying and classifying entanglement, as can be seen for example in ref. [10]. The possible partitions of the system may or may not be entangled, this being encoded by the dimension of the sheet we use in the Mayer-Vietoris subsequence. Of course, the classification given in this example is not complete, in the sense that it remains entanglement uncovered when looking only at a subsequence of the Mayer-Vietoris series and only at a particular dimension of the topological space. However, this article is the first to show that the long Mayer-Vietoris sequence has the capability to classify multipartite entanglement by means of a topological analysis, and provides the first potential extension of the ER-EPR result to multipartite entangled systems.


\begin{thebibliography}{99}
\bibitem{1} E. Schmidt, Math. Ann. 63, pag. 433 (1907)
\bibitem{2} C. H. Bennett, H. J. Bernstein, S. Popescu, B. Schumacher, Phys. Rev. A 53, 2046 (1996)
\bibitem{3} A. Acin, A. Andrianov, L. Costa, E. Jane, J. I. Latorre, R. Tarrach, Phys. Rev. Lett.  85, pag. 1560 (2000)
\bibitem{4} H. A. Carteret, A. Higuchi, A. Sudbery, J. Math. Phys. 41, 7932 (2000)
\bibitem{5} A. T. Patrascu, J. High Energ. Phys. (2017) 2017: 1 (2017)
\bibitem{6} J. Maldacena, L. Susskind, Fortsch. Phys. 61: pag. 781 (2013)
\bibitem{7} G. T. Whyburn, Amer. J. Math., 56, pag. 133 (1934)
\bibitem{8} E. G. Begle, Annals of Math. (2) 51, pag. 53 (1950)
\bibitem{9} J. M. Gary, Pacific J. of Math. 9, 4, pag. 1061 (1959)
\bibitem{10} S. Szalay, Phys. Rev. A 92, 042329 (2015)
\bibitem{11} J. S. Kim, G. Gour, B. C. Sanders, Contemp. Phys. 53 : 5 pag. 417 (2012)
\bibitem{12} L. Vietoris, "Uber die Homologiegruppen der Vereinigung zweier Komplexe", Monatshefte fur Mathematik 37, pag. 159-162 (1930)
\bibitem{13} N. Bao, J. Pollack, G. N. Remmen, Fortschr. Phys. 63, No. 11-12, pag. 705 (2015)
\bibitem{14} G. N. Remmen, N. Bao, J. Pollack, JHEP 07, 048 (2016)
\end{thebibliography}
\end{document}